# A Full Rank Pileup Deconvolution Scheme Suitable for Calorimeter Online Trigger Primitive Generation

Jinyuan Wu

***Abstract*— In this document, a pileup deconvolution scheme not relying on any mathematics guessing is presented. In high energy physics experiment, as the luminosity increases, pile-up issues on detectors such as calorimeters become non-negligible. Deconvolution approaches developed for data taken from DAQ systems are usually rank-deficient or underdetermined, having less equations than unknowns, even with the ADC values from multiple beam crossings are collected. These deconvolution approaches need mathematic pre-assumptions such as Sparse Representation. For online computation tasks such as for trigger primitive creation, signal availability is significantly different as in offline data analysis stage, and therefore, it is possible to use different (yet simpler) algorithms. In this situation, number of ADC values of the calorimeter outputs is the same as the number of beam crossings (or 4 times number of beam crossings, depending on the ADC sampling rate), and therefore, the number of equations can be arranged to be the same as number of unknowns. This way, a determined deconvolution scheme with a full-rank squared convolution (and deconvolution) matrix becomes possible. The robustness of deconvolution over long time windows is also studied in this paper.**

## I. Introduction

In modern high energy physics experiments, as the luminosity increases, pile-up issues become non-negligible. When a calorimeter ADC clock is synchronized with the accelerator RF, (for example in LHC, the ADC sampling clock could be either 40 or 160 MHz and locked with RF) the calorimeter readout data can be viewed as the convolution of the impulse response (IR) of the detector and the train of the event hits. The tail of the IR may extend several beam crossings (BC or BX) as shown in Fig. 1.

When several hits on the same calorimeter channel arrive within several BX, their signals superposition onto each other as shown above and the waveform may have peaks merged. To separate the contributions produces by close-by impulses from the readout signals, undoing the convolution is needed which is a deconvolution process.

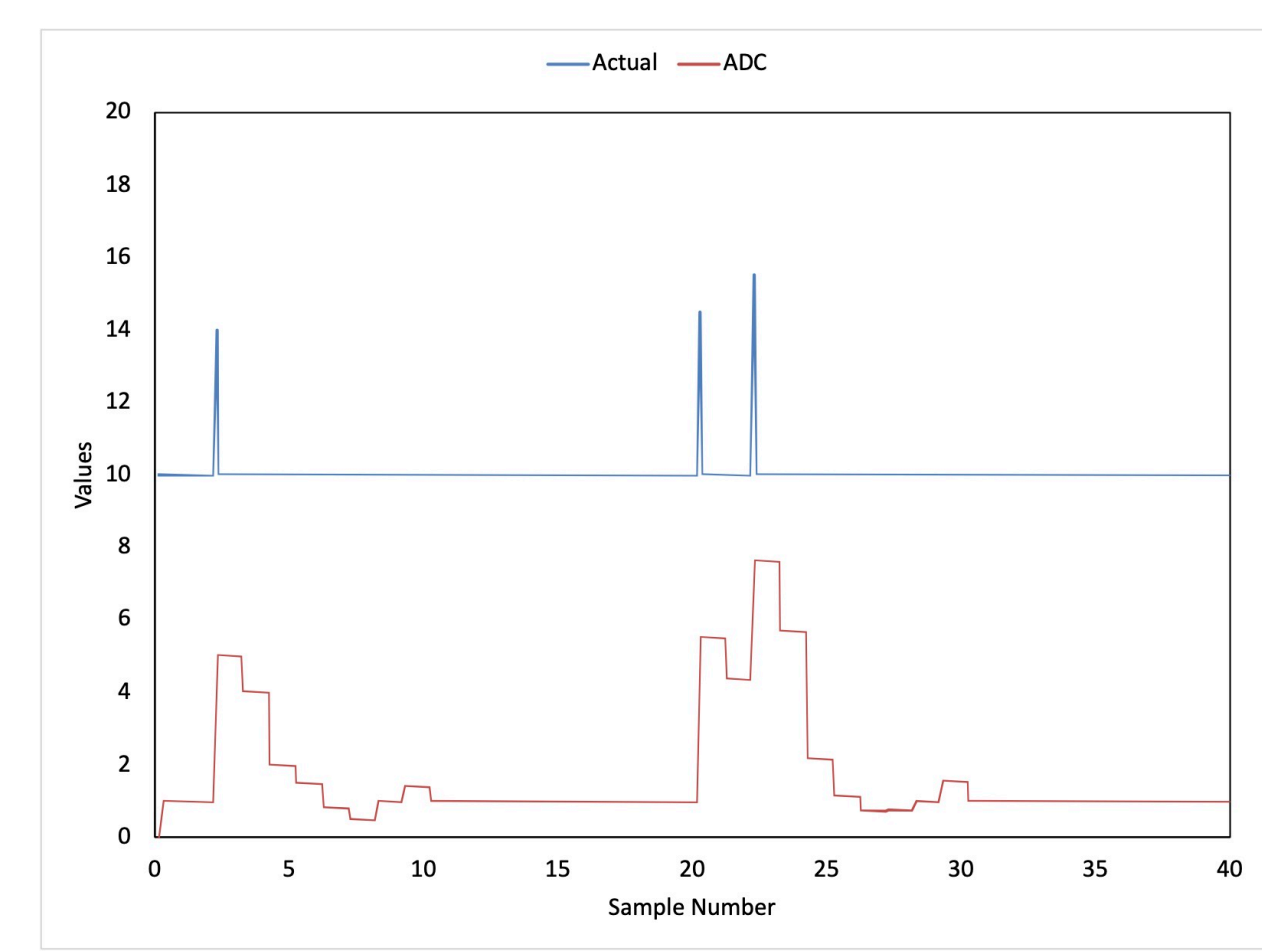

Fig. 1. Pile-up phenomenon.

For offline analysis, the data came from the DAQ system. The DAQ system will transmit and store the calorimeter data not only of the BX at the triggered events, but also several BX around as shown in Fig. 2.

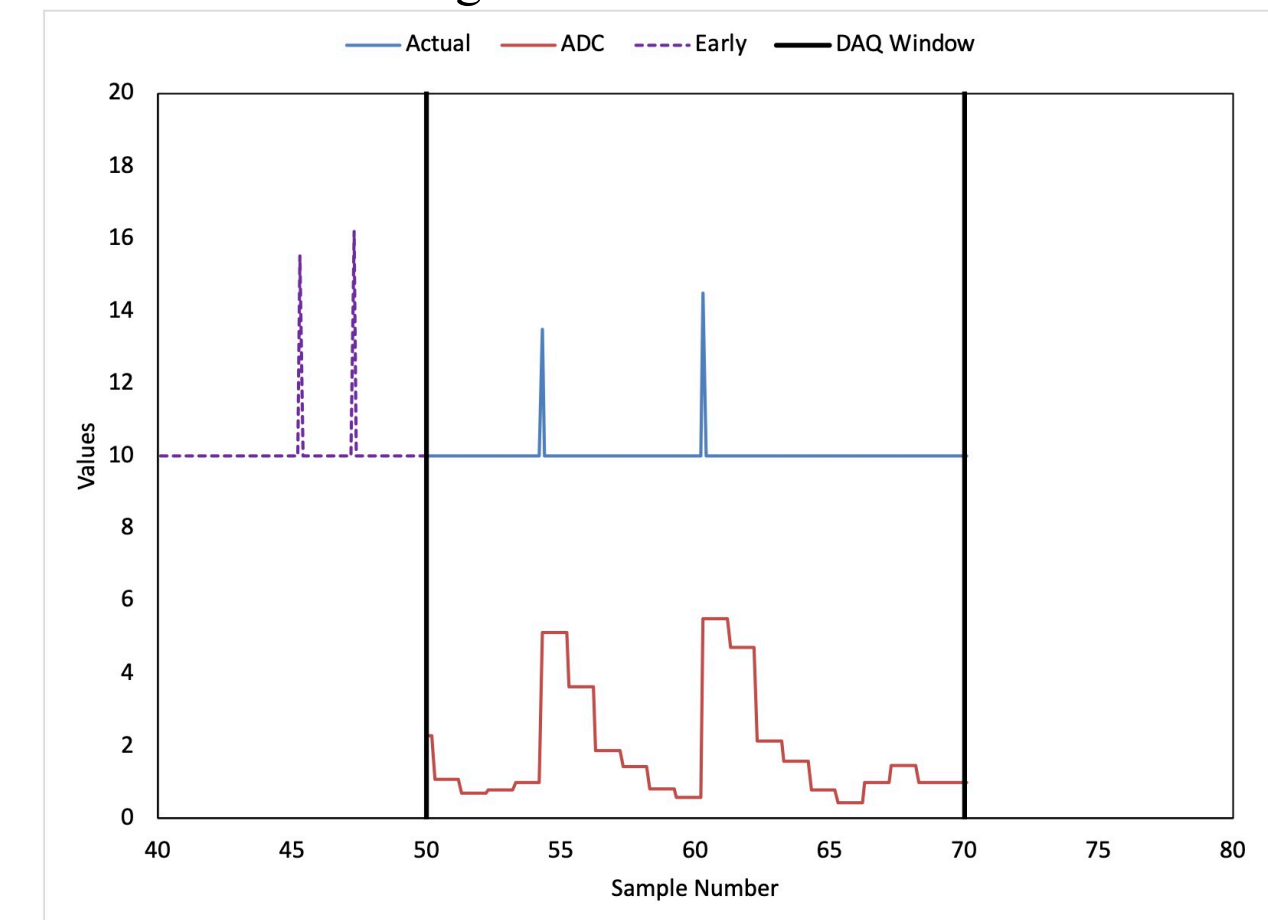

Fig. 2. Readout window of the DAQ system

If we can assume the calorimeter hits are all contained in the DAQ window, it is possible to fully recover values of the actual hits through deconvolution process for most practical detector

Manuscript submitted Oct. 17, 2025.

This manuscript has been authored by Fermi Forward Discovery Group, LLC under Contract No. 89243024CSC000002 with the U.S. Department of Energy, Office of Science, Office of High Energy Physics.

The author is with Fermi National Accelerator Laboratory, Batavia, IL 60510 USA (phone: 630-840-8911; fax: 630-840-2950; e-mail: jywu168@fnal.gov).

impulse responses. Assume the DAQ window has N data points (or BX if ADC sampling clock is the same as the BX frequency), the N amplitudes of the actual impulses are unknows to be solved, one is possible to solve N equations for N unknows (when the matrix of the systems of the equations is full column rank, which is true for most practical impulse response). But the early hits just before the DAQ window will also contribute to the data inside the window, therefore, there are more unknows than number of equations. The system of equations is underdetermined so that some pre-assumptions in mathematics must be applied to find the solution.

A well-developed approach is to use Sparse Representation that assume the measured waveform is produced by as few as possible impulses within both the DAQ window and the early range. References [1-5] have good reviews on this approach.

For online trigger primitive generation, the situation changes dramatically. Assuming the data processing FPGA is directly connected to the ADC, then all data samples are available for deconvolution. The system of the equations is not underdetermined anymore, so it is not necessary to use mathematics pre-assumptions such as Sparse Representation. In this document, the principle of the calorimeter deconvolution under online processing condition is first discussed in Section II. Results of several simple simulations are presented in Section III followed with conclusion in Section IV.

## II. Deconvolution for Online Applications

The online deconvolution requires all ADC readout values of the calorimeter, not just data in a timing window. This assumption is based on typical ADC-FPGA structure commonly seen in many HEP detector subsystems.

The calorimeter is considered as a linear time-invariant (LTI) system with finite impulse response (FIR) function h[i], where integer i is the time index, sample ID or beam crossing ID. The ADC output y[j] of the calorimeter is the convolution of the impulse response function and x[i], the train of the detector hits, which can be written into the following equation:

$$y[j] = \sum_{i=0}^{n} h[i]x[j-i] \qquad (1)$$

Note that the casualty condition has been applied, i.e., the output y[j] depends only on the impulses in the history, not in the future. We also assume the FIR has n+1 terms, and typically n < 10.

For general cases, it is not easy to find the solution of the equations giving in (1). In other word, finding unknowns of x[] from measurement values y[] is not simple.

Now we consider a simpler case with N+1 samples in a window with time index from k to (k + N). Define a vector $\mathbf{y}_0$ of calorimeter ADC outputs and a vector $\mathbf{x}_0$ of the input impulse, both with N+1 elements:

$$\mathbf{y_0} = \begin{pmatrix} y_0[k] \\ y_0[k+1] \\ y_0[k+2] \\ \vdots \\ y_0[k+N] \end{pmatrix}; \quad \mathbf{x_0} = \begin{pmatrix} x_0[k] \\ x_0[k+1] \\ x_0[k+2] \\ \vdots \\ x_0[k+N] \end{pmatrix} \qquad (2)$$

If all input impulses are within this N+1 points time window, no impulses are in the history before the k-th sample, then the convolution can be written in matrix format:

$$\mathbf{y_0} = \mathbf{H_0 x_0} \qquad (3)$$

The matrix $\mathbf{H}_0$ is a Toeplitz matrix with elements being the shifted version of the impulse response function h[i]:

$$\mathbf{H_0} = \begin{pmatrix} h[0] & 0 & 0 & 0 & 0 & 0 & 0 \\ h[1] & h[0] & 0 & 0 & 0 & 0 & 0 \\ \vdots & h[1] & h[0] & 0 & 0 & 0 & 0 \\ h[n] & \vdots & h[1] & h[0] & 0 & 0 & 0 \\ 0 & h[n] & \vdots & h[1] & h[0] & 0 & 0 \\ 0 & 0 & h[n] & \vdots & h[1] & h[0] & 0 \\ 0 & 0 & 0 & h[n] & \vdots & h[1] & h[0] \end{pmatrix} \qquad (4)$$

For most practical impulse response functions, $\mathbf{H}_0$ is full (column) rank so that its inverse $(\mathbf{H}_0)^{-1}$ exists. In this case, the original train of impulses x0 can be determined from the solutions of the system of the linear equations giving in (3):

$$\mathbf{x_0} = (\mathbf{H_0})^{-1}\mathbf{y_0} \qquad (5)$$

However, impulses in the history before the k-th sample do have contributions to the actual calorimeter ADC readout. So, the actual readout is the superposition of the contributions from the impulses within the time window and the ones in the history:

$$\mathbf{y} = \begin{pmatrix} y[k] \\ y[k+1] \\ y[k+2] \\ \vdots \\ y[k+N] \end{pmatrix} = \mathbf{y_1} + \mathbf{y_0} = \mathbf{H_1x_1} + \mathbf{H_0x_0} \qquad (6)$$

The perturbation term $\mathbf{y}_1$ has the same dimension as $\mathbf{y}$ and $\mathbf{y}_0$:

$$\mathbf{y_1} = \begin{pmatrix} y_1[k] \\ y_1[k+1] \\ y_1[k+2] \\ \vdots \\ y_1[k+N] \end{pmatrix} \qquad (7)$$

The train of impulses in the history is represented by vector $\mathbf{x}_1$, which has a dimension of n, the length of impulse response (minus 1):

$$\mathbf{x_1} = \begin{pmatrix} x[k-n] \\ \vdots \\ x[k-2] \\ x[k-1] \end{pmatrix} \qquad (8)$$

The matrix of convolution for the perturbation in the history is also a Toeplitz matrix of with elements being the shifted version of the impulse response function h[n]:

$$\mathbf{H_1} = \begin{pmatrix} h[n] & \cdots & h[3] & h[2] & h[1] \\ 0 & h[n] & \vdots & h[3] & h[2] \\ \vdots & 0 & h[n] & \vdots & h[3] \\ \vdots & \vdots & 0 & h[n] & \vdots \\ 0 & \vdots & \vdots & 0 & h[n] \\ 0 & 0 & \vdots & \vdots & 0 \\ 0 & 0 & 0 & 0 & 0 \end{pmatrix} \qquad (9)$$

The matrix $\mathbf{H}_1$ is a "tall" one with n columns and N rows. In actual computation, only the top n elements need to be calculated. The only reason why 0's are padded in the lower rows is to match the dimension of other matrix in Equation (6).

When $\mathbf{y}$ and $\mathbf{x}_1$ are known, the unknowns of $\mathbf{x}_0$ can be calculated with the following formula:

$$\mathbf{x_0} = (\mathbf{H_0})^{-1}(\mathbf{y_0} - \mathbf{H_1x_1}) \qquad (10)$$

The Equation (6) can be visualized in Fig. 3.

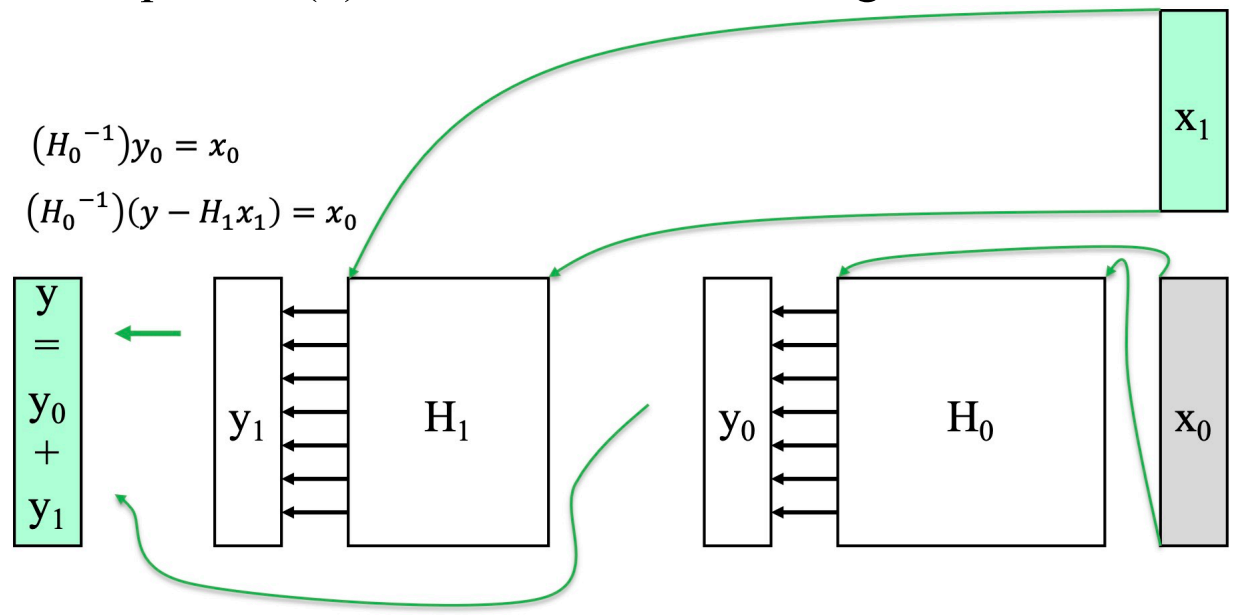

Fig. 3. The deconvolution scheme

The deconvolution process starts at a sufficiently long beam gap, such as a gap due to accelerator revolution. In this situation, it is known there is no calorimeter hits due to beam and therefore $\mathbf{x}_1$ in Equation (10) is 0. The train of impulses in $\mathbf{x}_0$ is calculated and lower portion of $\mathbf{x}_0$ becomes $\mathbf{x}_1$ for calculation in the next timing window. The above process repeats so that the entire train of the impulses is calculated.

Note that although not all impulses will be output and stored in DAQ system, they should still be calculated since the algorithm is a recursive one.

To reduce accumulation of the computation errors, one may force set $\mathbf{x}_0 = \mathbf{0}$ if the calculated results show that there is no significant calorimeter hit in the time window other than random noise.

## III. Simulation Results

To validate the deconvolution scheme described in previous section, a simple simulation is studied. The simulation condition and output results are described in this section.

The impulse response vector of the calorimeter in our simulation has finite length. The values of the terms in the impulse response are shown in Table I.

Table I
Impulse response of the calorimeter in the simulation

| h[0] | h[1] | h[2] | h[3] | h[4] | h[5] | h[6] | h[7] |
|---|---|---|---|---|---|---|---|
| 1 | 0.75 | 0.25 | 0.125 | -0.05 | -0.125 | 0 | 0.1 |

The impulse response in this example is essentially a fast decay with some ringing near the tail.

The primary convolution matrix, i.e., the matrix $\mathbf{H}_0$ discussed in the previous section has columns essentially the shifted versions of the impulse response vector. The values of the matrix elements are shown in Table II.

Table II
Elements of the primary convolution matrix

| 1 | 0 | 0 | 0 | 0 | 0 | 0 | 0 | 0 | 0 |
|---|---|---|---|---|---|---|---|---|---|
| 0.75 | 1 | 0 | 0 | 0 | 0 | 0 | 0 | 0 | 0 |
| 0.25 | 0.75 | 1 | 0 | 0 | 0 | 0 | 0 | 0 | 0 |
| 0.125 | 0.25 | 0.75 | 1 | 0 | 0 | 0 | 0 | 0 | 0 |
| -0.05 | 0.125 | 0.25 | 0.75 | 1 | 0 | 0 | 0 | 0 | 0 |
| -0.125 | -0.05 | 0.125 | 0.25 | 0.75 | 1 | 0 | 0 | 0 | 0 |
| 0 | -0.125 | -0.05 | 0.125 | 0.25 | 0.75 | 1 | 0 | 0 | 0 |
| 0.1 | 0 | -0.125 | -0.05 | 0.125 | 0.25 | 0.75 | 1 | 0 | 0 |
| 0 | 0.1 | 0 | -0.125 | -0.05 | 0.125 | 0.25 | 0.75 | 1 | 0 |
| 0 | 0 | 0.1 | 0 | -0.125 | -0.05 | 0.125 | 0.25 | 0.75 | 1 |

Due to causality, the primary convolution matrix is a lower triangular one.

The history perturbation matrix, i.e., the matrix $\mathbf{H}_1$ discussed in previous section is shown in Table III.

Table III
Elements of the history perturbation matrix

| 0.1 | 0 | -0.125 | -0.05 | 0.125 | 0.25 | 0.75 |
|---|---|---|---|---|---|---|
| 0 | 0.1 | 0 | -0.125 | -0.05 | 0.125 | 0.25 |
| 0 | 0 | 0.1 | 0 | -0.125 | -0.05 | 0.125 |
| 0 | 0 | 0 | 0.1 | 0 | -0.125 | -0.05 |
| 0 | 0 | 0 | 0 | 0.1 | 0 | -0.125 |
| 0 | 0 | 0 | 0 | 0 | 0.1 | 0 |
| 0 | 0 | 0 | 0 | 0 | 0 | 0.1 |
| 0 | 0 | 0 | 0 | 0 | 0 | 0 |
| 0 | 0 | 0 | 0 | 0 | 0 | 0 |
| 0 | 0 | 0 | 0 | 0 | 0 | 0 |

Note that the matrix is a "tall" one with 0's padded in the lower rows. Disregarding the lower all 0 rows, the matrix is a upper triangular one.

The inverse $(\mathbf{H}_0)^{-1}$ of the primary convolution matrix should be pre-calculated for the deconvolution algorithm described in Equation (10), and its elements are shown in Table IV.

Table IV
Elements of the inverse matrix of the primary convolution matrix

| 1 | 0 | 0 | 0 | 0 | 0 | 0 | 0 | 0 | 0 |
|---|---|---|---|---|---|---|---|---|---|
| -0.75 | 1 | 0 | 0 | 0 | 0 | 0 | 0 | 0 | 0 |
| 0.3125 | -0.75 | 1 | 0 | 0 | 0 | 0 | 0 | 0 | 0 |
| -0.17188 | 0.3125 | -0.75 | 1 | 0 | 0 | 0 | 0 | 0 | 0 |
| 0.194531 | -0.17188 | 0.3125 | -0.75 | 1 | 0 | 0 | 0 | 0 | 0 |
| -0.05449 | 0.194531 | -0.17188 | 0.3125 | -0.75 | 1 | 0 | 0 | 0 | 0 |
| -0.0644 | -0.05449 | 0.194531 | -0.17188 | 0.3125 | -0.75 | 1 | 0 | 0 | 0 |
| -0.03192 | -0.0644 | -0.05449 | 0.194531 | -0.17188 | 0.3125 | -0.75 | 1 | 0 | 0 |
| 0.110096 | -0.03192 | -0.0644 | -0.05449 | 0.194531 | -0.17188 | 0.3125 | -0.75 | 1 | 0 |
| -0.0762 | 0.110096 | -0.03192 | -0.0644 | -0.05449 | 0.194531 | -0.17188 | 0.3125 | -0.75 | 1 |

In the simulation, a set of random impulses are generated with pre-defined occupancy (10% in the example shown below). The impulse responses for the hits are calculated and a random noise with pre-defined level (~9% p-p of the maximum amplitude of the impulse in the example shown below) is added as the readout data of the ADC.

The deconvolution process giving in the Equation (10) is run using the data of ADC with noise. The results are shown in Fig. 4.

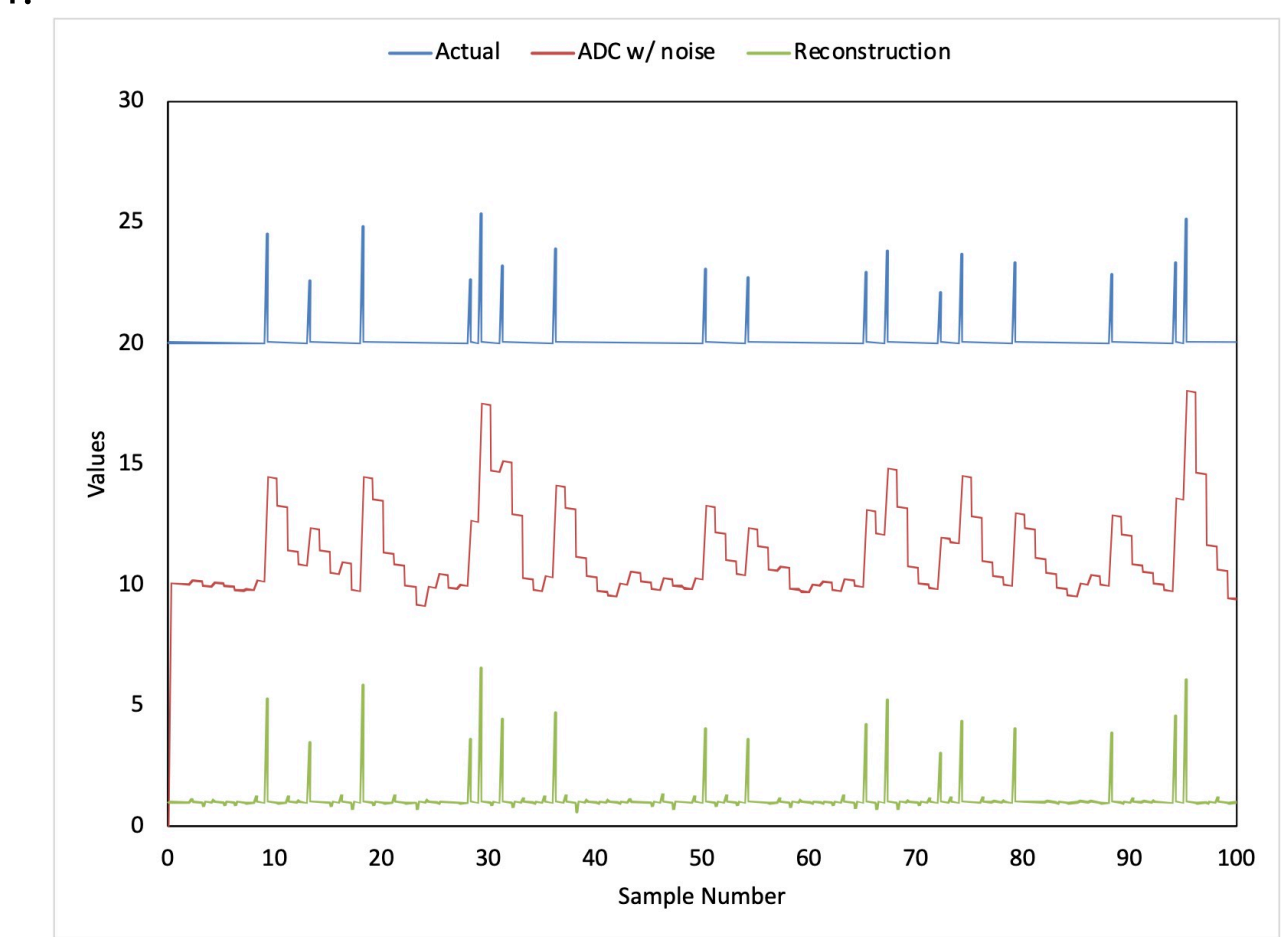

Fig. 4. Deconvolution results

One can see that the actual impulses are recovered correctly, even for the ones in close-by beam crossings with peaks smeared together. The noise added to the ADC outputs creates some ghost impulses, both positive and negative, but the ghosts are sufficiently small comparing to the actual ones.

Since the deconvolution process is a recursive one, meaning there could be long data dependency in the long history, noise accumulation is an issue one must address. To study the possible noise accumulation, the simulation is run over a relatively long timing window. The results are shown in Fig. 5.

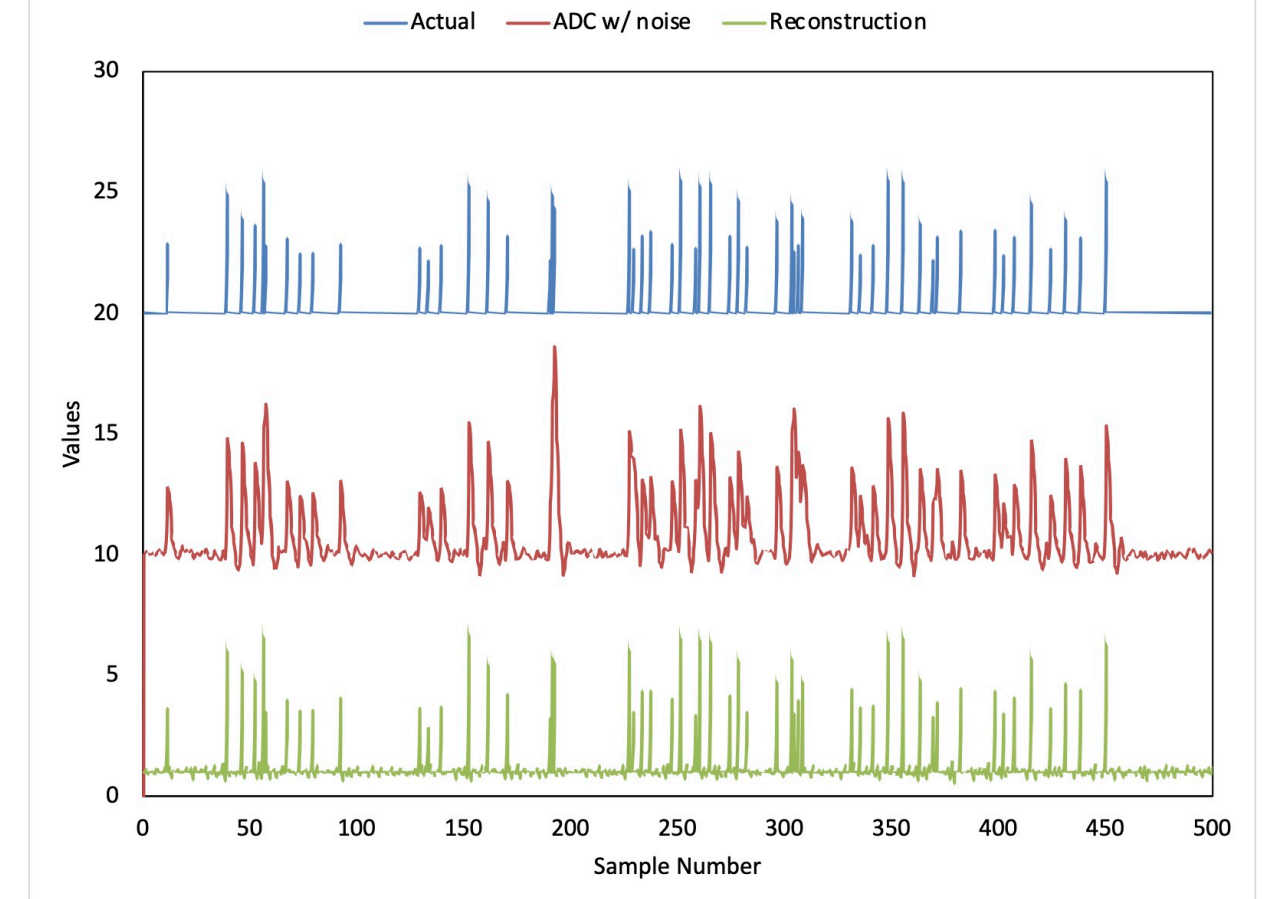

Fig. 5. Simulation over a long timing window

From the output results, one can see that under the condition in the simulation, there is no visible noise accumulation. Once the decay in the impulse response is fast enough, it is anticipated that the gain of the noise across the recursive iteration is much smaller than 1 which will suppress the noise accumulation.

## IV. Discussions

The deconvolution scheme described in this document is based purely on the measured ADC values. The mathematics pre-assumptions such as sparse representation used in mainstream approaches are not needed in our scheme giving that all ADC data are known in the FPGA for online computation. The actual FPGA implementation will be the next step toward the goal for online applications.